\documentclass[floatfix,aps,pre,twocolumn]{revtex4}
\usepackage{graphics}
\usepackage[pdftex]{graphicx}               
\begin{document}
\addtolength{\topmargin}{0.5in}



\title{Jamming of 3D Prolate Granular Materials}


\author{K.~Desmond}
\author{Scott~V.~Franklin}
\email[]{svfsps@rit.edu}
\homepage[]{http://piggy.rit.edu/franklin/}
\affiliation{Dept. of Physics, Rochester Institute of Technology}

\date{\today}

\begin{abstract}
We have found that the ability of long thin rods to jam into a
solid-like state in response to a local perturbation depends upon both
the particle aspect ratio and the container size.  The dynamic phase
diagram in this parameter space reveals a broad transition region
separating granular stick-slip and solid-like behavior.  In this
transition region the pile displays both solid and stick-slip
behavior.  We measure the force on a small object pulled through the
pile, and find the fluctuation spectra to have power law tails with an
exponent characteristic of the region.  The exponent varies from
$\beta=-2$ in the stick-slip region to $\beta=-1$ in the solid region.
These values reflect the different origins --- granular rearrangements
vs. dry friction --- of the fluctuations.  Finally, the packing
fraction shows only a slight dependence on container size, but depends
on aspect ratio in a manner predicted by a mean field theory and
implies an aspect-ratio independent contact number of $\langle c
\rangle = 5.25 \pm 0.03$.
\end{abstract}

\maketitle
\section{introduction}

Anyone scooping nails from a bin at the hardware store or shoveling
hay with a pitchfork has noticed the ability of large aspect-ratio
($\alpha\equiv L/D\gg1$) granular materials to act as a solid.  This
state can exist at low packing fractions, raising the possibility of
lightweight, rigid building materials.  Such materials have practical
applications as proppant in oil recovery or conducting networks in
quantum dot solar cells.  Here we present the first
experimental investigation into the dependence of this state on
particle aspect ratio and container size.

The jamming of round and low aspect-ratio ($<5$) materials has
attracted a great deal of recent attention.  A canonical example of
granular jamming is flow through hoppers, where arches at the exit can
support the entire pile \cite{To, To2, To3, Zuriguel05}.  This jammed
state, while capable of supporting large constant forces, is not
robust when subjected to sharp taps.  Indeed, in one set of
experiments \cite{Zuriguel05} a jet of air was sufficient to
reintroduce flow.

Jamming is also invoked to explain the erratic drag force on an object
moving through a granular medium \cite{RAlbert, Albert2}.  The object
comes to rest when it encounters a connected network of particles that
is incapable of rearranging; only when the applied force is large
enough to break through this network does the motion begin again.  The
particles in this network are only a small fraction of the total
grains in the pile, those comprising the force chains \cite{Ohern,
Coppersmith} that terminate on the container walls.

The situation is remarkably different for high aspect-ratio ($>10$)
granular materials \cite{Stokely}.  Philipse \cite{Philipse1} noted
that particles of aspect ratio larger than 35 emerged as a solid plug
when poured.  This plug maintained the shape of the original container
even in face of external disturbances.  This jammed state can
therefore be considered both global and robust, the connected network
of particles containing a significant fraction of the pile that are
incapable of moving around one another.  How this solid-like state
compares with the jammed state of ordinary granular materials has not
yet been explored.

Recent work has investigated the mean drag force exerted by ordinary
granular materials on an intruder.  Zhou et al. found \cite{Zhou04,
Zhou05}, for example, that mean drag scales linearly with pressure for
both mono- and poly-disperse materials.  Geng and
Behringer\cite{Geng2005} studied the force fluctuations exerted by a
two-dimensional packing of disks and found that the power spectra had
a power law tail with exponent $-2$, as in three-dimensional systems.
Hill et al.\cite{Hill05} and Stone et al.\cite{Stone04, StoneII04}
have all looked at plates pushed slowly down into beds of sand, with
interesting behaviors as the plate approaches the bottom of the
container and a temporal evolution in the granular bed with successive
plungings.  Finally, Bratberg et al.\cite{Bratberg2005} simulated the
quasi-static flow of rigid, frictional disks pushed upward against
gravity through a narrow pipe.  They observed a transition in the flow
when the intruder speed was large.

\section{Setup and Packing Fraction}

In our experiments, the force on a small ball pulled upward through a
pile of prolate granular materials is measured.  Particles are cut
from acrylic rods of diameter $d=1/8''$ or $d=1/16''$ to varying
lengths; aspect ratios vary from $4$ to $48$.  The particles are
dropped into cylinders with diameters ranging from $D=1''$ to $D=4''$.
A dimensionless container diameter is formed by dividing by the
particle length ($\tilde D \equiv D/L$); the smallest $\tilde D$ is
0.67, the largest 12.  At the bottom of the pile sits a small
($d=0.25''$) metal ball connected to string that runs up through the
pile.  We have confirmed that packings formed around the string have
the same packing fraction as those formed in the string's absence.
The string is wound around the shaft of a motor that turns with
constant angular speed.  A force sensor measures the tension in the
string to within $0.5\%$ at kHz resolution.  For the range of forces
in our experiments, the string and force sensor act Hookian with
spring constant $k=1600$ N/m.  The motion of the ball can be erratic;
when the resistance of the pile exceeds that applied by the string the
ball will stop moving until the applied force, increasing as the
string is wound, is again large enough to induce motion.

\begin{figure}[ht]
\includegraphics[angle=-90,scale=0.8]{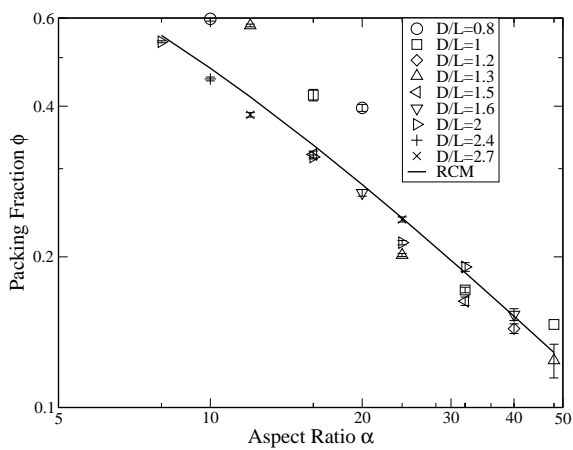}
\caption{Log-log plot of packing fraction $\phi$ as a function of
  particle aspect ratio $\alpha$.  Data are grouped by constant
  dimensionless container diameter $\tilde D \equiv D/L$.  With the
  exception of a few points at low $\tilde D$ (where significant
  particle alignment occurs), the data are well-fit by a mean-field
  theory with only one free parameter.\label{pack}}
\end{figure}

Particles are simply dropped into the cylinder.  ``Fluffing'' the pile
with air, a technique used to prepare ordinary granular materials in a
low density state, does not significantly decrease the packing
fraction of rods \cite{Lumay04}.  The packing fraction is plotted
(log-log) versus aspect ratio in Fig.~\ref{pack}, with different
symbols indicating different dimensionless container sizes.  The data
from different size containers are reasonably close to one another,
except for a few points at the low $\tilde D$.  At these values the
particles tend to align with the sidewalls, introducing significant
orientational correlation and higher packing fractions.

To gain some idea of the randomness of the initial packing we compare
the packing fraction with a mean-field theory the {\it Random Contact
Model} (RCM) \cite{Philipse1}.  The basic assumption of this model is
that contacts between particles are uncorrelated, a significant
difference from the correlated contacts that result in ordered packing
of, for example, disks or rods.  If the contacts are uncorrelated,
then the packing can be thought of as a collection of independent
pairs of rods in contact.

A key finding of Philipse is that the average number of contacts
$\langle c \rangle$ scales linearly with with the particle
concentration.  To show this, Philipse begins with the fraction of
orientations made inaccessible to a particle by the existence of
another particle.  This is a function of the center-of-mass separation
of the two particles; when two rods are close together, they must be
more nearly aligned to prevent overlapping.  Earlier work on the
packing of rods in two dimensions \cite{Stokely} calculated the
functional dependence of the excluded fraction of orientations $f_{\rm
ex}(\vec r)$ analytically and found that, while both experiment and
simulations contained long-range correlations, the functional form was
universal over several orders of magnitude in aspect ratio.  The
average number of contacts is then proportional to the product of the
excluded fraction of orientations and the local number density:
$$\langle c \rangle = {1 \over 2}\int f_{\rm ex}(\rho, \vec r) \rho
(\vec r) {\rm d}\vec r.$$
The proportionality factor of 1/2 avoids double counting of contacts.  

The mean-field approximation replaces the local density $\rho(\vec r)$
with the average pile density $\bar \rho$; the remaining integral over
the excluded orientations can then be identified with the average
excluded volume $v_{\rm ex}$, defined \cite{Onsager} as the volume
denied to particle $j$ by the condition that it not overlap with
particle $i$.  The average contact number then scales linearly with
packing concentrations and the number density of a pile is
$$\rho = {2 \langle c \rangle \over v_{\rm ex}}.$$

The excluded volume for cylinders of length $L$ and thickness $D$ is
\cite{Onsager}
\begin{eqnarray*}
v_{\rm ex} & = &(\pi/2)L^2d + {\pi (\pi + 3) \over 4}Ld^2 + (\pi^2
/8)d^3\\
& = &L^3 \left [ {\pi \over 2}\alpha + {\pi (\pi + 3) \over 4}\alpha^2 +
(\pi^2/8)\alpha^3\right ].
\end{eqnarray*}
and, since the particle volume is $v_p=(\pi/4)d^2L$ and $\phi=\rho
v_p$, the packing fraction is
\begin{eqnarray}
\phi & = & {2 \langle c \rangle (\pi/4)d^2L \over L^3 \left [
(\pi/2)\alpha + {\pi (\pi + 3) \over 4}\alpha^2 + (\pi^2/8)\alpha^3
\right ]} \\ & = & {4 \pi \langle c \rangle \alpha \over 4 \pi+ 2 \pi
(\pi + 3) \alpha + \pi^2\alpha^2 }\ .\label{RCM}
\end{eqnarray}
The predicted packing fraction $\phi(\alpha)$ has one single free
parameter, the average contact number $\langle c \rangle$.  We (and
others \cite{Lumay04}) noticed that significant orientational order exists when the
cylinder diameter $D$ is less than 1 particle length $L$.  This
violates the main assumption of the model, and so we should not expect
good agreement in this region.  We therefore attempt to fit the
prediction from Eq.~\ref{RCM} with all data taken in cylinders where
$D/L>1$.  The resulting line is shown in Fig.~\ref{pack}. For large
aspect ratios Eq.~\ref{RCM} reduces to $\phi \sim \langle c
\rangle/\alpha$, a scaling that was noticed previously in
2-\cite{Stokely} and 3-dimensions \cite{Philipse1}.  The curvature in
the data in Fig.~\ref{pack} shows that this simple limit does not
apply; the full expression of Eq.~\ref{RCM} (solid line) agrees quite
well with the data.  From the fit we find $\langle c \rangle = 5.25\pm
0.03$, which agrees (within reported error) with the data reported by
Philipse \cite{Philipse1}.  The constant contact number implies a
significant screening effect, since the contact number could in
principle scale with aspect ratio.  A similar result for two
dimensional piles was found in earlier simulations\cite{Stokely}.

The good agreement between experiment and a model that assumes an
isotropic distribution of particle angle supports the claim that
particles are initially randomly oriented.  The discrepancy at small
container size is understandable, as the orientational correlation
induced by the narrow cylinder violates the fundamental assumption of
the RCM.

\begin{figure}[ht]
\begin{center}
\includegraphics[angle=-90,scale=0.7]{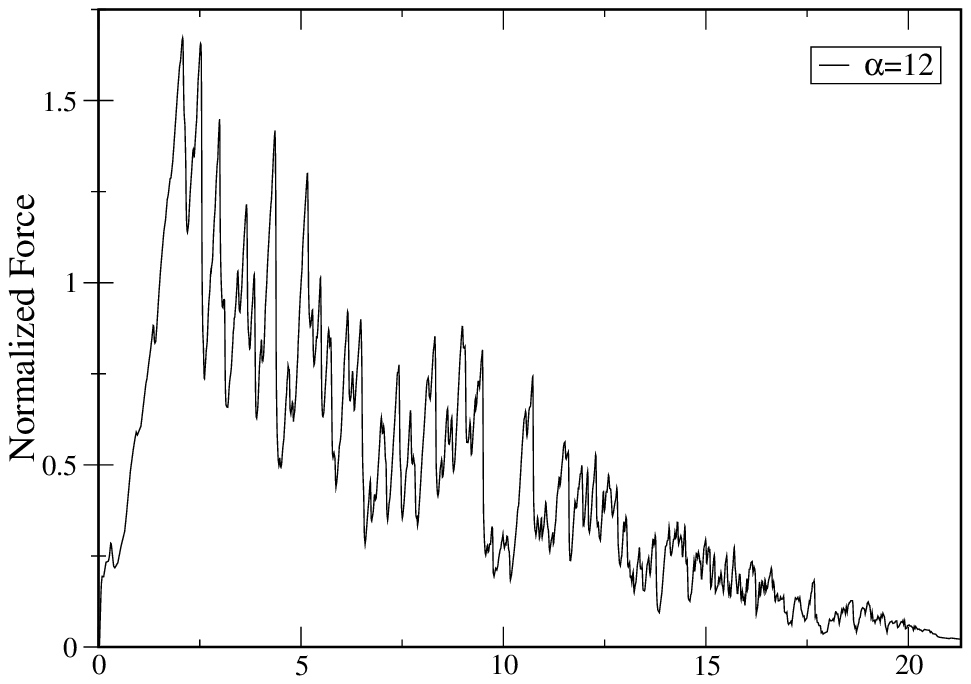}

\includegraphics[angle=-90,scale=0.7]{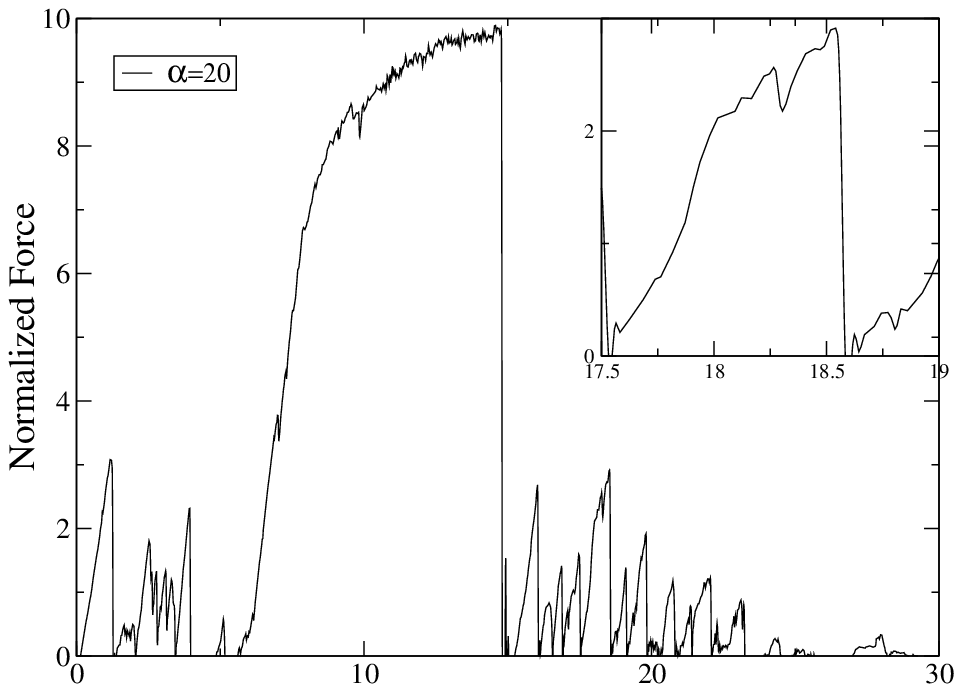}

\includegraphics[angle=-90,scale=0.7]{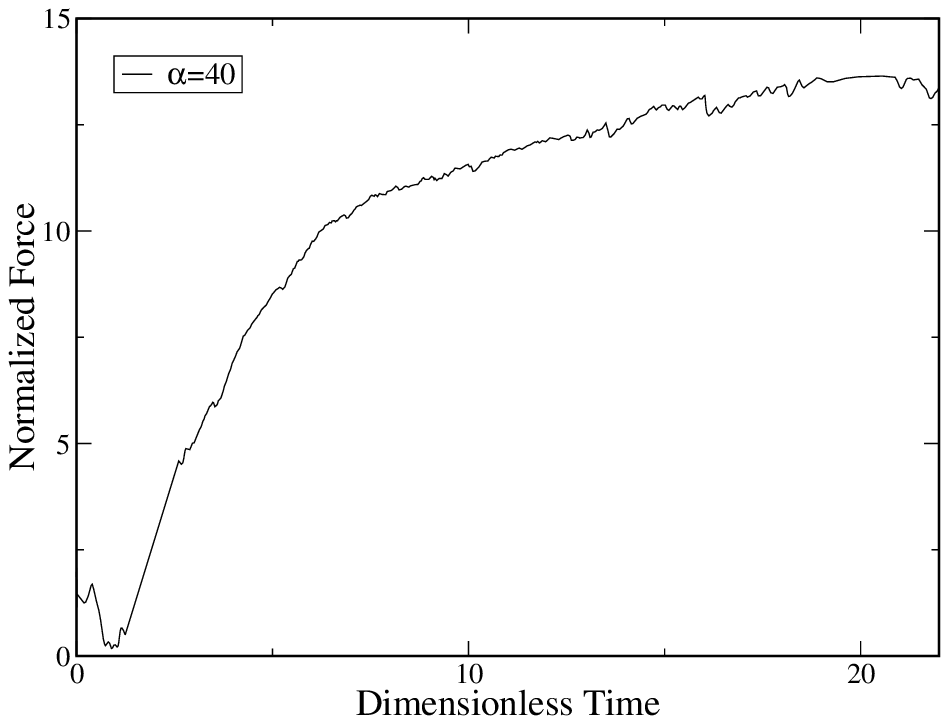}
\caption{Force vs. time data for a ball dragged through a pile of rods
  of aspect ratio 12 (top), 20 (middle), and 40 (bottom), all in a 2''
  tube.  The low aspect ratio particles show the stick-slip behavior
  common in ordinary granular materials.  The large aspect-ratio
  particles, however, act as a single solid body, with small
  fluctuations characteristic of dry friction.  Intermediate aspect
  ratios show both behaviors in a single experiment on both large and
  small (inset) time scales.
  \label{forces}}
\end{center}
\end{figure}
\section{Force Scaling in Different Regions}

When the particle aspect ratio is low, the pile responds with local
rearrangements and the drag force on an intruding object has a
random-sawtooth appearance (Fig.~\ref{forces}(top)).  The force
increases linearly, indicating that the ball is at rest, before
rapidly decreasing, indicating a burst of motion.  The ball is brought
to rest and the cycle repeats.  Throughout the experiment, the bulk
pile is at rest.  While individual particles near the ball are moving,
there is no collective pile motion.  As the aspect ratio is increased,
however, the pile exhibits a qualitative change to solid-like
behavior.  With the exception of a few stray particles, all particles
are lifted upward, with no observed relative motion between particles.
The resulting force vs. time diagram is shown in
Fig.~\ref{forces}(bottom). Forces are normalized by the total pile
weight, and so Fig.~\ref{forces} shows that the force required to move
the pile can be many times the actual pile weight.  This is a
consequence of force chains \cite{Ohern, Coppersmith} that terminate
on the container walls.  The ball is pushing upward on the particles,
but the lateral deflections of the force chains translate this into a
force normal to container walls.  This normal force can be quite
large, resulting in large frictional forces between the walls and the
pile.

\begin{figure}[ht]
\includegraphics[angle=-90,scale=0.85]{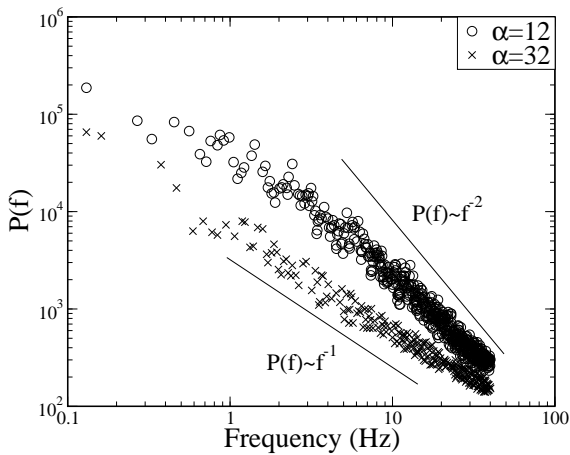}
\caption{\label{fluct}Power spectra for force fluctuations from piles
  exhibiting granular ($\circ$) and solid (x) behavior.  Both show
  power law tails, with the different exponents indicating a different
  mechanism for the fluctuations.  The $1/f$ and and $1/f^2$ decay are
  consistent with previous work on, respectively, dry friction and
  localized granular rearrangements.  Both data sets taken in a D=2''
  diameter tube.}
\end{figure}

For intermediate aspect ratios, the pile displays characteristics of
both the granular and solid states.  Figure~\ref{forces}(middle) shows
data from one such run.  The pile first behaves much like that of
smaller aspect ratio particles, although a close examination of the
force (inset in Fig.~\ref{forces}(middle)) reveals small plateaus
indicating small amounts of solid-body behavior.  Then, with the ball
in the middle of the pile, the entire pile jams and all particles
above the ball are lifted.  Corresponding to the plateaus in the force
is the collective motion of the entire pile.  The pile is visually
observed to move as a solid body for a brief period of time.  The
breakup of the solid state is brought about by a collapse of the
particles around the ball.  As the pile is constantly rubbed against
the sidewall (and, indeed, shows force fluctuations characteristic of
dry friction) the length of time spent in the solid state is some
indication of the pile's stability.

A Fourier transform of the force vs. time data in the granular and
solid regions (Fig.~\ref{fluct}) shows that both spectra show power
law tails, albeit with different exponents indicating a different
fluctuation origin.  Data from the granular region decay as $P(f)\sim
f^{-2}$, consistent with earlier experiments on granular materials
\cite{Albert, Miller}.  Coupled with the visual observation of no
macroscopic pile motion, this supports connecting the force serrations
with local particle rearrangements.  The forces in the solid region,
however, decay as $P(f)\sim f^{-1}$.  This is consistent with
experimental work on dry friction \cite{Feder91}, and so we infer that
the pile is not moving steadily upward, but in fact sticking on the
side walls.

\begin{figure}[ht]
\includegraphics[angle=-90,scale=.85]{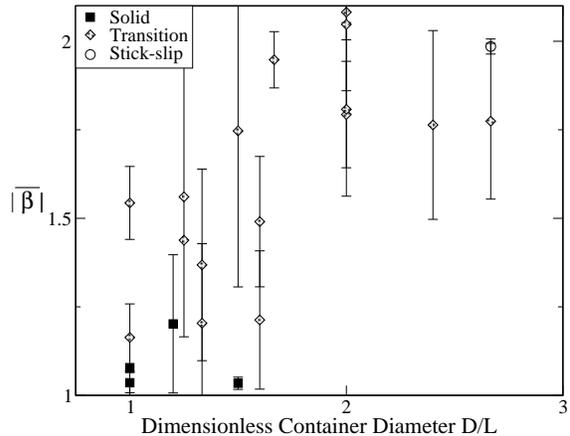}
\caption{\label{beta} Exponents of the spectra tails for experimental
  runs of varying aspect ratios plotted vs. normalized container
  diameter $\tilde D = D/L$.  Filled symbols are aspect ratios larger
  than 20, which display solid-like motion in containers less than 1.5
  particle lengths across.}
\end{figure}

The exponent of the spectrum tail is a reliable indicator of the
pile's behavior.  Shown in Fig.~\ref{beta} is the exponent magnitude
$\mid \beta \mid $ as a function of normalized tube diameter $\tilde
D$.  The data are grouped in three sets corresponding to the three
different behaviors: stick-slip ($\circ$), transition ($\diamond$),
and solid-like (\raisebox{3pt}{\framebox[6pt][t]{\ }}).  Lying off the
graph's scale are three additional stick-slip data points (each
representing an average of several runs) at $\tilde D = 5, 8, \& 12$;
all are characterized by an exponent of magnitude of just under 2 with
relatively small uncertainties.  Data exhibiting transitional behavior
show much larger uncertainties, indicating the statistically
fluctuating nature of the behavior.  While all force spectra from
these experiments show the plateaus characteristic of the transition
region, the relative amount of time spent on a plateau (as opposed to
stick-slip) varies wildly from run to run.  Data in the solid region
have exponents close to one with small uncertainties (with the
exception of data at $\tilde D = 1.2$, which we suspect was very close
to the transition region).

\section{Dynamic Phase Diagram}

\begin{figure}[ht]
\includegraphics[angle=-90,scale=.85]{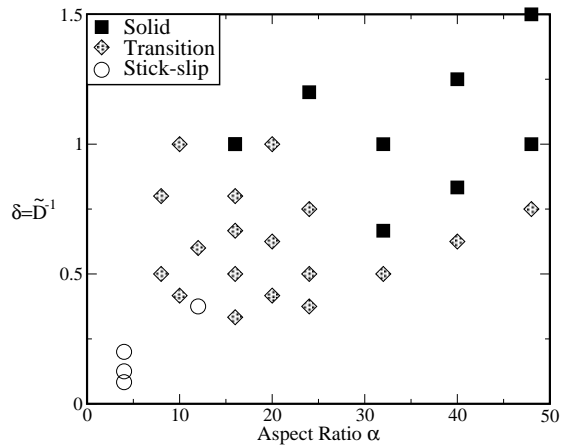}
\caption{\label{phase}Phases exhibited by granular piles as a function
of two control parameters --- the aspect ratio $\alpha$ and the
inverse container diameter $\delta \equiv \tilde D^{-1}$.  Smaller aspect ratio
particles show the stick-slip behavior of granular materials, while
larger aspect ratio particles act as a solid body when the container
is small enough.}
\end{figure}

To better understand the transition from granular to solid-like
motion, we have mapped out the dynamic phase diagram of the behavior
as a function of the two control parameters --- aspect ratio $\alpha$
and container diameter $\tilde D$.  It is in fact more revealing to
use the inverse container diameter $1/\tilde D$.  This is because when
the particle length is increased but tube diameter held fixed both the
aspect ratio and inverse diameter ($\delta\equiv L/D$) increase
linearly, and so a sequence of experiments in which particle length is
increased shows up as straight lines in $\delta - \alpha$ space.  This
is shown in Fig.~\ref{phase}.

As one would expect, when the aspect ratio is very small, the pile
behaves in a canonically granular manner.  Nevertheless, the signature
characteristics of the transition region --- plateaus in the force
data and a visual observation of collective motion --- are seen in
particles with aspect ratios as low as 8 when confined to cylinders
whose diameter is twice the particle length.  As the container
diameter is further reduced, and the ability of particles to move
around one another further constrained, we expect solid-body motion to
occur even for the small aspect ratios (upper left corner of the phase
diagram), but we did not explore this region.

We did not investigate the dependence of the transition region on
pulling speed.  There are two different mechanisms through which the
pulling speed can become significant: elastic unloading (particularly
significant in a stiff machine) and particle rearrangements.  We show
here that we are well below the critical velocities at which either of
these become important.  In a theoretical investigation of frictional
disks pushed upward through a vertical pipe, Bratberg et
al.\cite{Bratberg2005} found a transition that depended on the
relative value of two time scales, $t_k=({m \over k})^{1/2}$ ($m$ the
pile weight, $k$ the spring constant of the pulling mechanism)
corresponding to the elastic unloading and $t_g={v \over g}$ the time
for a particle to accelerate due to gravity to velocity $v$.  An
order-of-magnitude estimation of these values for our system is
$m\approx 1$ kg, $v=0.01$ m/s, and $k=1600$ N/m, and therefore
$${t_k \over t_g}={\sqrt{m \over k} \over v/g} \approx {10^{-2} \ {\rm
s} \over 10^{-3} \ {\rm s}}\approx 10.$$ Using these numbers, we would
expect a significant influence of pushing velocity to occur at a
characteristic velocity of about $v_c = \sqrt{m \over k}g \approx 24$
cm/s, twenty times our actual pulling speed.  Therefore, our pulling
``spring'' is soft enough, and our velocities slow enough, so that the
drop in force is not significant.

If the particle rearrangements are governed solely by gravitational
forces, then the ratio of the time it takes a particle to fall under
gravity a distance equal to its length $L$ $t_r=\sqrt{L \over g}$ to
the time it takes the intruder to move that distance $t_i=L/v$ is
important.  The ratio of these times, using our approximate
parameters, is then
$${t_i \over t_r} = {\sqrt{gL} \over v} \approx 50,$$ leading to a
critical velocity of $v_c=\sqrt{gL}\approx 50$ cm/s, about 50 times
faster than the current pulling speed.  Thus, the intruder moves a
negligible distance while the grains themselves are re-orienting.  The
above analyses imply that we are in the quasi-static region of
behavior and will not see significant changes in the behavior unless
the pulling speed is increased by at least an order of magnitude.

Philipse observed that particles with aspect ratios larger than 35 did
not flow when poured from a bucket, but rather emerged as a single,
solid plug.  The explanation given was a geometric entanglement of the
rods. Interestingly, however, we do not observe the solid behavior in
particles with an aspect ratio of 48 when poured into the large
cylinder.  It may be that, in response to the localized disturbance of
a small intruder, the pile can make small rearrangements necessary to
allow the intruder to pass through while still maintaining an overall
rigid structure.  An important next step in this work will be to
investigate the effect of intruder size on phase behavior.  For small
intruders, we imagine we are probing details of the spaces (voids) in
the pile while larger intruders probe the pile's rigidity.  This would
imply a critical length scale related to particle length and width,
consistent with our observation that the particle aspect ratio alone
does not completely determine the pile's behavior.

\section{Conclusions}

We have observed three qualitatively different types of behavior in
large aspect-ratio granular materials in response to a local
disturbance: canonically granular stick-slip, solid-body like motion,
and a transition region that is a combination of the two.  The phase
space of this behavior has been mapped out as a function of two
control parameters, the particle aspect ratio and container diameter.
We have found that even low aspect-ratio particles can exhibit
temporary solid-body motion that is characteristic of transitional
behavior.  Surprisingly, the larger aspect ratio particles do not
behave as solid bodies when the container is large enough.  The
granular and solid states both show force fluctuations with spectra
that have power-law tails, although the exponents are
characteristically different.

We have also investigated the packing fraction as a function of these
two control parameters.  While there is a slight dependence on
container size, we find a distinct dependence on aspect ratio that
agrees quite well with the mean-field Random Contact Model.  This is
perhaps surprising, as we observe some orientational ordering that
violates a main assumption of this model.

\begin{acknowledgments}
This research was supported by an award from the Research
Corporations; Scott Franklin is a Cottrell Scholar of Research
Corporation.  Acknowledgment is also made to the Donors of the
American Chemical Society Petroleum Research Fund for support of this
research.
\end{acknowledgments}


\bibliography{../../Reference_lists/references}
\end{document}